# Closed-Loop Load Model Identification Using Small Disturbance Data

Shangyuan Li
*Department of Electrical Engineering*
Zhejiang University
Hangzhou, China
429042268@qq.com

Li Feng
*Department of Dispatch*
State Grid Chongqing Electric Power Company
Chongqing, China
287428203@qq.com

Deqiang Gan
*Department of Electrical Engineering*
Zhejiang University
Hangzhou, China
dgan@zju.edu.cn

Zhen Wang
*Department of Electrical Engineering*
Zhejiang University
Hangzhou, China
eezwang@ieee.org

Wei Bao
*Department of Electrical Engineering*
Zhejiang University
Hangzhou, China
wbao@mrpower.cn

Hao Xu
*Department of Electrical Engineering*
Zhejiang University
Hangzhou, China
xuhao0729@qq.com

*Abstract*—Load model identification using small disturbance data is studied. It is proved that the individual load to be identified and the rest of the system forms a closed-loop system. Then, the impacts of disturbances entering the feedforward channel (internal disturbance) and feedback channel (external disturbance) on relationship between load inputs and outputs are examined analytically. It is found out that relationship between load inputs and outputs is not determined by load itself (feedforward transfer function) only, but also related with equivalent network matrix (feedback transfer function). Thus, load identification is closed loop identification essentially and the impact of closed loop identification cannot be neglected when using small disturbance data to identify load parameters. Closed loop load model identification can be solved by prediction error method (PEM). Implementation of PEM based on a Kalman filtering formulation is detailed. Identification results using simulated data demonstrates the correctness and significance of theoretical analysis.

*Keywords—load identification, closed loop identification, prediction error method, Kalman filtering*

## I. Introduction

The importance of load modelling to power system stability evaluation and control design has been well recognized in power community [1]. It is reported that, using different load models, one can produce different even contradictory simulation results [2, 3].

Due to the wide-spread deployment of wide-area measurement system (WAMS), the measurement based method has attracted much attention in recent years. The method establishes a composite load model firstly. Then, the method searches for a set of optimal parameters so that the output of composite load model best fits with actual output [4]. The related research focuses on three issues, they are: how to determine the structure of the composite load model [4]; how to reduce computation during identification process [5, 6]; how to search for these parameters [7, 8].

However, the current practice in measurement-based load modelling is to use large disturbance data, as the result of a three-phase-to-ground fault or a move in tap changer position, to identify load parameters. Unfortunately, large disturbance data are of very limited availability. As a result, using small disturbance (say, random load variations) data to model load becomes tempting because small disturbance data for any period is readily available. This is the motivation of this research.

This paper is organized as follows. Section 2 and 3 show that load identification actually bears a closed-loop nature, a fact that has detrimental impact in small-disturbance based identification, while in large-disturbance based identification the consequence of this fact is not that significant. In section 4, it is shown that the closed loop identification problem can be solved by a prediction error method (PEM) if some mild requirements are met. Section 5 details the implementation of PEM using a Kalman filtering strategy. In section 6, simulations identification results are reported to demonstrate the potential of the introduced findings.

## II. The Closed-Loop Nature of Load Identification

As is well-known, a power system is constantly excited by small disturbances, like random load variations. These random disturbances may occur inside the load to be identified, or distant away from the load, see Fig. 1. Hereinafter, disturbances inside the load to be identified are named internal disturbances, like disturbance in mechanical torque of an induction motor load. While, disturbances outside are named external disturbances, like random variations in other loads, network operations, variations in generator outputs. Since the system is subject to small disturbances, it is reasonable to assume that the system dynamics is linear.

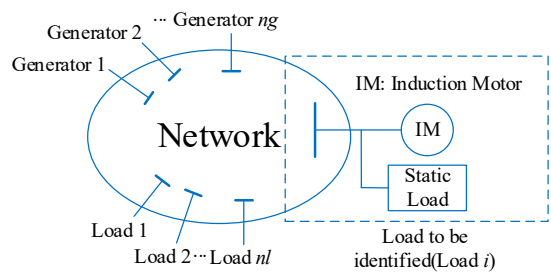

Fig. 1 Schematic diagram of a power system

### A. The load model

Assume there are $nl+ng$ buses in system, the first $ng$ buses are generators buses, and the latter $nl$ buses are load buses. For the objective here is to identify the model of the load in bus $i$, thus dynamic characteristic of load model in bus $i$ should considered accurately and other loads in system can be simplified for ease of explanation. In this paper, the other loads in system are regarded as constant impedances. The load to be identified is a usual combination of a static part model and a dynamic third-order induction motor model [9].

The load model to be identified in bus $i$ can take the following form without loss of generality:

$$\Delta \mathbf{I}_{Li}(s) = \mathbf{G}_{Li}(s)\Delta \mathbf{V}_{Li}(s) + \mathbf{H}_i(s)\xi_i(s) \quad (1)$$

where $\Delta \mathbf{I}_{Li}(s)=[\Delta I_{xi}(s)\ \Delta I_{yi}(s)]^T$ is the load current which can be viewed as the output of the load to be identified, $\Delta \mathbf{V}_{Li}(s)=[\Delta V_{xi}(s)\ \Delta V_{yi}(s)]^T$ is bus voltage which can be viewed as the input of the load. Subscript $xi$ and $yi$ denotes real part and imaginary part of variable in bus $i$ respectively. Assume for simplicity there is only one internal disturbance denoted by $\xi_i(s)$ in load $i$. Accordingly, $\mathbf{G}_{Li}(s)\in 2\times 2$ is the load model to be identified, and $\mathbf{H}_i(s)\in 2\times 1$ denotes an unknown transfer function from internal disturbance to outputs.

*B. The generator model*

The $ng$ linearized generator models can be described by the following model:

$$\begin{aligned} s\Delta \mathbf{x}_g(s) &= \mathbf{A}_g \Delta \mathbf{x}_g(s) + \mathbf{B}_g \Delta \mathbf{V}_g(s) \\ \Delta \mathbf{I}_g(s) &= \mathbf{C}_g \Delta \mathbf{x}_g(s) + \mathbf{D}_g \Delta \mathbf{V}_g(s) \end{aligned} \quad (2)$$

where $\Delta \mathbf{x}_g(s)$ denotes state variable of all generators. $\Delta \mathbf{V}_g(s)$ denotes inputs of all generators. $\Delta \mathbf{I}_g(s)$ denotes outputs of all generators. In the above equation, $\mathbf{A}_g, \mathbf{B}_g, \mathbf{C}_g, \mathbf{D}_g$ are diagonal matrixes. Details about the generator models can be found, say, in [1].

*C. The network model*

The linearized network model can be described as follows:

$$\Delta \mathbf{I}(s) = \mathbf{Y} \Delta \mathbf{V}(s) \quad (3)$$

where $\mathbf{Y}\in 2(nl+ng)\times 2(nl+ng)$ is the network admittance matrix, $\Delta \mathbf{V}\in 2(nl+ng)\times 1$ is the network voltage vector, $\Delta \mathbf{I}\in 2(nl+ng)\times 1$ is the current vector. The disturbances in other loads can be modelled as random current injections into the network. Without loss of generality, suppose there is only one external disturbance at load $h$, let it be denoted by $\xi_h(s)$. It is clear now that $\Delta \mathbf{I}$ and $\Delta \mathbf{V}$ in equation (3) takes the following form: the first $2ng$ variables in $\Delta \mathbf{I}$ and $\Delta \mathbf{V}$ are generators variables, the last $2nl$ terms in $\Delta \mathbf{I}$ and $\Delta \mathbf{V}$ are load variables. Because $\mathbf{Y}$ has included the equivalent impedance of all loads except the load in bus $i$, most variables in last $2nl$ terms in $\Delta \mathbf{I}$ are zeros except the load to be identified at bus $i$ and external disturbance at load $h$.

*D. The complete closed loop system*

State space model as shown in (2) can be re-written in transfer function form:

$$\Delta \mathbf{I}_g(s) = [\mathbf{C}_g(s\mathbf{I}-\mathbf{A}_g)^{-1}\mathbf{B}_g + \mathbf{D}_g]\Delta \mathbf{V}_g(s) \quad (4)$$

Substituting (4) into (3), we can eliminate the first $2ng$ variables in $\Delta \mathbf{I}(s)$ to obtain:

$$\Delta \overline{\mathbf{I}}(s) = \overline{\mathbf{Y}}(s) \Delta \mathbf{V}(s) \quad (5)$$

where $\overline{\mathbf{Y}}$ has included $\mathbf{C}_g(s\mathbf{I}-\mathbf{A}_g)^{-1}\mathbf{B}_g + \mathbf{D}_g$, the transfer function, and $\Delta \mathbf{I}(s)$ after elimination is denoted by $\Delta \overline{\mathbf{I}}(s)$.

Eliminating the rest of variables except $\Delta I_{xi}(s), \Delta I_{yi}(s), \Delta V_{xi}(s)\ \Delta V_{yi}(s), \xi_h(s)$, equation (5) is further simplified into:

$$\begin{bmatrix} \Delta I_{xi}(s) \\ \Delta I_{yi}(s) \end{bmatrix} = \mathbf{K}(s) \begin{bmatrix} \Delta V_{xi}(s) \\ \Delta V_{yi}(s) \end{bmatrix} + \mathbf{K}_h(s)\xi_h(s) \quad (6)$$

where $\mathbf{K}(s)$ is the equivalent network matrix and $\mathbf{K}_h(s)$ is transfer function from external disturbance to load outputs. For analytical convenience, equation (6) is transformed into a compact form as:

$$\Delta \mathbf{I}_{Li}(s) = \mathbf{K}(s)\Delta \mathbf{V}_{Li}(s) + \mathbf{K}_h(s)\xi_h(s) \quad (7)$$

It follows that:

$$\Delta \mathbf{V}_{Li}(s) = \mathbf{K}^{-1}(s)\Delta \mathbf{I}_{Li}(s) - \mathbf{K}^{-1}(s)\mathbf{K}_h(s)\xi_h(s) \quad (8)$$

Based on (8) and (1), it is understood that the load to be identified and the rest of the system forms a closed-loop system as shown in Fig. 2. As can be seen from the figure, there are two excitation sources in the closed loop system: internal disturbance $\xi_i(s)$ and external disturbance $\xi_h(s)$ respectively. Internal disturbance and external disturbance enters the feedforward and feedback channel respectively. The impacts of internal disturbance and external disturbance on load model identification are different and will be studied in section 3.

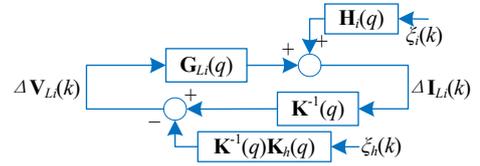

Fig. 2 The complete closed loop system

### III. THE PITFALL OF CLOSED LOOP IDENTIFICATION

When using large disturbance data to perform load model identification, the fact that the system is in a closed-loop setup has been completely neglected. As will be shown shortly, this has minimal impact on the accuracy of load modelling. However, this is not the case when using small disturbance data to perform load model identification. The reason of this controversy is described in this section.

Notice that $\Delta \mathbf{I}_{Li}(s)$ and $\Delta \mathbf{V}_{Li}(s)$ are determined according to superposition principle of linear systems:

$$\begin{aligned} \Delta \mathbf{I}_{Li}(s) &= \Delta \mathbf{I}_{\xi i}(s) + \Delta \mathbf{I}_{\xi h}(s) \\ \Delta \mathbf{V}_{Li}(s) &= \Delta \mathbf{V}_{\xi i}(s) + \Delta \mathbf{V}_{\xi h}(s) \end{aligned} \quad (9)$$

where $\Delta \mathbf{I}_{\xi i}(s), \Delta \mathbf{V}_{\xi i}(s)$ are responses caused by internal disturbance $\xi_i(s)$, and $\Delta \mathbf{I}_{\xi h}(s), \Delta \mathbf{V}_{\xi h}(s)$ are responses caused by external disturbance $\xi_h(s)$. From (1), (7) and (8), it is straightforward to obtain the following results:

$$\begin{aligned} \Delta \mathbf{V}_{\xi i}(s) &= \left[\mathbf{K}(s)-\mathbf{G}_{Li}(s)\right]^{-1}\mathbf{H}_i(s)\xi_i(s) \\ \Delta \mathbf{I}_{\xi i}(s) &= \mathbf{K}(s)\left[\mathbf{K}(s)-\mathbf{G}_{Li}(s)\right]^{-1}\mathbf{H}_i(s)\xi_i(s) \\ \mathbf{V}_{\xi h}(s) &= \left[\mathbf{G}_{Li}(s)-\mathbf{K}(s)\right]^{-1}\mathbf{K}_h(s)\xi_h(s) \\ \Delta \mathbf{I}_{\xi h}(s) &= \mathbf{G}_{Li}(s)\left[\mathbf{G}_{Li}(s)-\mathbf{K}(s)\right]^{-1}\mathbf{K}_h(s)\xi_h(s) \end{aligned} \quad (10)$$

After some simple manipulations, it follows:

$$\Delta \mathbf{I}_{Li}(s) = \left[\mathbf{K}(\mathbf{K}-\mathbf{G}_{Li})^{-1}\mathbf{H}_i\xi_i + \mathbf{G}_{Li}(\mathbf{G}_{Li}-\mathbf{K})^{-1}\mathbf{K}_h\xi_j\right] \times \\ \left[(\mathbf{K}-\mathbf{G}_{Li})^{-1}\mathbf{H}_i\xi_i + (\mathbf{G}_{Li}-\mathbf{K})^{-1}\mathbf{K}_h\xi_j\right]^{-1} \times \Delta \mathbf{V}_{Li}(s) \quad (11)$$

where Laplace operator $s$ in (11) is omitted. The above closed-form formula reveals that the true relationship between load outputs $\Delta \mathbf{I}_{Li}(s)$ and load inputs $\Delta \mathbf{V}_{Li}(s)$ is a weighted average of feedforward $\mathbf{G}_{Li}(s)$ and feedback $\mathbf{K}^{-1}(s)$, with the disturbance magnitudes of $\xi_i(s)$ and $\xi_h(s)$ being the weights. If $\xi_i(s)$ does not exist, the relationship between $\Delta\mathbf{I}_{Li}(s)$ and $\Delta\mathbf{V}_{Li}(s)$ is determined by $\mathbf{G}_{Li}(s)$ even though feedback channel exists, that is $\Delta\mathbf{I}_{Li}(s) = \mathbf{G}_{Li}(s)\Delta\mathbf{V}_{Li}(s)$.

When $\xi_i(s)$ presents and $\xi_h(s)$ does not, the relationship is simplified into:

$$\Delta\mathbf{I}_{Li}(s) = \mathbf{K}(s)\Delta\mathbf{V}_{Li}(s) \quad (12)$$

This is the most unfavourable situation that can occur.

In a typical load model identification setup, both internal and external disturbance present. However, if the external disturbance $\xi_h(s)$ is far greater than internal disturbance $\xi_i(s)$, then (11) is approximately simplified into the desired form:

$$\Delta\mathbf{I}_{Li}(s) \approx \mathbf{G}_{Li}(s)\Delta\mathbf{V}_{Li}(s) \quad (13)$$

This is the case when identifying load models using large disturbance data. When identifying load models using small disturbance data, the situation is more involved, and this is the theme of subsequent sections.

## IV. Direct Load Model Identification Using PEM

In preceding section, it is shown that load model identification is actually in a closed-loop setup, and the input-output of the load does not necessarily reflect the true dynamics of the load. To cope with this difficulty, several methods judiciously designed for closed-loop identification has been reported [14]. Prediction error method (PEM) is chosen in this work for load model identification because it gives statistically efficient estimate under rather mild conditions. Besides, PEM is particularly suitable for load identification since it demands input-output data only, as in small disturbance data based load model identification, only input-output data is available. In what follows we briefly review the method and exam the suitability of it. PEM proceeds with a discrete system model and discrete data. A general discrete MIMO system is shown in Fig. 3.

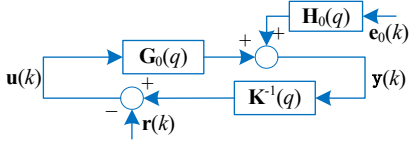

Fig. 3 MIMO system with feedback

Without loss of generality, a MIMO model to be identified is shown in (14) below:

$$\mathbf{y}(k) = \mathbf{G}_0(q,\vartheta_G)\mathbf{u}(k) + \mathbf{H}_0(q,\vartheta_H)\mathbf{e}_0(k) \quad (14)$$

where $q$ represents the shift operator, $(\vartheta_G, \vartheta_H)$ are the model parameters to be sought ($\vartheta$ for short). Now, $\mathbf{G}(q,\vartheta_G)$ is used to represent MIMO model obtained by PEM, $\mathbf{H}(q,\vartheta_H)$ is used to represent internal disturbance model, $\mathbf{e}_0(k)$ is the unknown internal disturbance vector. It is assumed that $\mathbf{y}$, $\mathbf{u}$, and $\mathbf{e}_0$ are random signals that are quasi-stationary and ergodic, and the covariance matrix of $\mathbf{e}_0(k)$ is $\Lambda_0$.

The idea of PEM is to estimate the unknown parameters $(\vartheta_G, \vartheta_H)$ simultaneously, using input $\mathbf{u}(k)$ and output $\mathbf{y}(k)$ directly, regardless the presence of feedback channel. Assume $N$ data records with sampling frequency $T_s$ are generated by the system:

$$\mathbf{u}(k), \mathbf{y}(k), k = 1,...,N \quad (15)$$

Let $\hat{\mathbf{y}}(k|k-1)$ be the conditional expectation of $\mathbf{y}(k)$, then

$$\hat{\mathbf{y}}(k|k-1) = \mathbf{H}^{-1}(q,\vartheta_H)\mathbf{G}(q,\vartheta_G)\mathbf{u}(k) - [1-\mathbf{H}^{-1}(q,\vartheta_H)]\mathbf{y}(k)$$

[13]. It follows that the one-step prediction error vector for the model is

$$\varepsilon(k|\vartheta) = \mathbf{H}^{-1}(q,\vartheta_H)[\mathbf{y}(k) - \mathbf{G}(q,\vartheta_G)\mathbf{u}(k)] \quad (16)$$

PEM searches for unknown parameters $(\vartheta_G, \vartheta_H)$ to minimize the following loss function $V(\vartheta)$:

$$\min_{\vartheta} V(\vartheta) = \frac{1}{N}\sum_{k=1}^{N}\frac{1}{2}\varepsilon^T(k|\vartheta)\varepsilon(k|\vartheta) \quad (17)$$

The above time-domain optimization formulation has a frequency-domain equivalence, as $N \to \infty$ [14]:

$$\min_{\vartheta} V(\vartheta) = \frac{1}{2\pi}\int_{-\pi}^{\pi} tr(\Phi_\varepsilon)d\omega \quad (18)$$

where $\Phi_\varepsilon$ is the spectrum of the prediction error $\varepsilon(k,\vartheta)$, tr denotes the trace of the matrix. To proceed the analysis, we need to fix some terminology. We will denote $\mathbf{G}(q,\vartheta_G)$ as $\mathbf{G}_\vartheta$, $\mathbf{H}(q,\vartheta_G)$ as $\mathbf{H}_\vartheta$ for short. Introduce the signal $\chi_0 = [\mathbf{u}^T(k)\ \mathbf{e}_0^T(k)]$, the spectrum of it is:

$$\Phi_{\chi_0}(\omega) = \begin{bmatrix} \Phi_u(\omega) & \Phi_{ue}(\omega) \\ \Phi_{eu}(\omega) & \Lambda_0 \end{bmatrix} \quad (19)$$

where $\Phi_u(\omega)$ is the spectrum of $\mathbf{u}(k)$, $\Phi_{ue}(\omega)$ is cross spectrum between $\mathbf{u}(k)$ and $\mathbf{e}_0(k)$, and $\Phi_{eu}(\omega)$ is cross spectrum between $\mathbf{e}_0(k)$ and $\mathbf{u}(k)$. Let $\hat{\vartheta}_N$ be the solution of the error minimization problem, superscript * denote complex conjugate transpose. In the PEM regime, a fundamental result is as follows [14]: Suppose the data spectrum matrix $\Phi_{\chi_0}(\omega)$ be positive definite ($\Phi_{\chi_0} > 0$) for all frequencies, then, with probability 1 as $N \to \infty$,

$$\hat{\vartheta}_N \to \arg\min_{\vartheta} \int_{-\pi}^{\pi} tr\{[(\mathbf{G}_0 - \mathbf{G}_\vartheta)\ (\mathbf{H}_0 - \mathbf{H}_\vartheta)]\Phi_{\chi_0} \\ \times \begin{bmatrix}(\mathbf{G}_0 - \mathbf{G}_\vartheta)^* \\ (\mathbf{H}_0 - \mathbf{H}_\vartheta)^*\end{bmatrix}(\mathbf{H}_\vartheta\mathbf{H}_\vartheta^*)^{-1}\}d\omega \quad (20)$$

The condition $\Phi_{\chi_0} > 0$ is termed the experiment is *informative*. Since $\Phi_{\chi_0}$ is not easily computed as it requires the information of $\mathbf{e}_0(k)$, an alternative test is used in this work. Let $\mathbf{z}(k) = [\mathbf{u}^T(k)\ \mathbf{y}^T(k)]$, then if $\Phi_z(\omega) > 0$, the experiment is informative. It is interesting that the above is only a sufficient condition, the readers are referred to [15] for more discussions in this line of research.

Another basic pre-requisite of PEM is that the input data, $\mathbf{u}(k)$, $k=0,1,...,N$, should be *persistently exciting*. The sequence $\mathbf{u}(k)$ is persistently exciting of order $n$ if and only if the following matrix is non-singular [10]

$$\mathbf{R}^u = \begin{bmatrix} R_u(0) & \cdots & R_u(n-1) \\ \vdots & \ddots & \vdots \\ R_u(n-1) & \cdots & R_u(0) \end{bmatrix} \quad (21)$$

where $R_u(\tau)$ is auto correlation matrix. The physical meaning of persistently exciting of order $n$ is that the input data can be used to estimate model of order lower than $n$.

In the presence of feedback, we have:

$$\mathbf{u}(k) = \mathbf{r}(k) - \mathbf{K}^{-1}(q)\mathbf{y}(k) \quad (22)$$

where $\mathbf{K}^{-1}(q)$ is the transfer function matrix of the feedback and $\mathbf{r}(k)$ is the external disturbance. If $\mathbf{r}(k)=0$, and $\mathbf{K}^{-1}(q)$ is a constant matrix, then it is easily shown that the identification is *not* feasible [13]. A general conclusion in the context of closed-loop identification is that, in order to find accurate identification solution, the feedback $\mathbf{K}^{-1}(q)$ should be of sufficiently high-order [13]. This is fortunately the case in this study.

To study the bias of estimation, let us introduce $\mathbf{\Phi}_e^r$:

$$\mathbf{\Phi}_e^r = \mathbf{\Lambda}_0 - \mathbf{\Phi}_{eu}\mathbf{\Phi}_u\mathbf{\Phi}_{ue} \quad (23)$$

The following result characterizes the bias distribution of closed-loop identification [14]. Suppose the input data $\mathbf{u}(k)$ is persistently exciting, then, with probability 1 as $N \to \infty$,

$$\hat{\vartheta}_N \to \arg\min_{\vartheta} \int_{-\pi}^{\pi} tr\{[(\mathbf{G}_0 + \mathbf{B}_\vartheta - \mathbf{G}_\vartheta)\mathbf{\Phi}_u(\mathbf{G}_0 + \mathbf{B}_\vartheta - \mathbf{G}_\vartheta)^* \\ + (\mathbf{H}_0 - \mathbf{H}_\vartheta)\mathbf{\Phi}_e^r(\mathbf{H}_0 - \mathbf{H}_\vartheta)^*][\mathbf{H}_\vartheta\mathbf{H}_\vartheta^*)^{-1}\}d\omega \quad (24)$$

where

$$\mathbf{B}_\vartheta = (\mathbf{H}_0 - \mathbf{H}_\vartheta)\mathbf{\Phi}_{eu}\mathbf{\Phi}_u^{-1} \quad (25)$$

It becomes clear that, if $\mathbf{B}_\vartheta = 0, \mathbf{\Phi}_u > 0$, as $V(\vartheta)$ vanishes, one obtains the desired estimate $\mathbf{G}_\vartheta = \mathbf{G}_0$. This can happen either in the open-loop case where $\mathbf{\Phi}_{eu} = 0$, or in the situation where the internal disturbance model is independently identified so $(\mathbf{H}_0 - \mathbf{H}_\vartheta) = 0$.

If, however, the internal disturbance model is not independently identified in a closed-loop setup, then one would find a biased estimate:

$$\mathbf{G}_\vartheta = \mathbf{G}_0 + \mathbf{B}_\vartheta \quad (26)$$

The bias can be measured in terms of an upper bound on $\bar{\sigma}(\mathbf{B}_\vartheta)$ (the maximum singular value of $\mathbf{B}_\vartheta$):

$$\bar{\sigma}(\mathbf{B}_\vartheta) \le \bar{\sigma}(\mathbf{H}_0 - \mathbf{H}_\vartheta)\sqrt{\frac{\bar{\sigma}(\mathbf{\Lambda}_0)}{\underline{\sigma}(\mathbf{\Phi}_u)}}\sqrt{\frac{\bar{\sigma}(\mathbf{\Phi}_u^e)}{\underline{\sigma}(\mathbf{\Phi}_u)}}$$

Here $\underline{\sigma}(\mathbf{\Phi}_u)$ represents the minimum singular value of $\mathbf{\Phi}_u$, and $\mathbf{\Phi}_u^e$ denotes the part in $\mathbf{\Phi}_u$ coming from $\mathbf{e}_0(k)$.

The implications of the above result is self-evident. Under the condition that the experiment is informative and input data is persistently exciting, the bias would tend to be small if either of the following holds:

1. $\bar{\sigma}(\mathbf{H}_0 - \mathbf{H}_\vartheta)$ is small, it means that the internal disturbance model is good;

2. $\bar{\sigma}(\mathbf{\Phi}_u^e)/\underline{\sigma}(\mathbf{\Phi}_u)$ is small, it means that the feedback internal disturbance contribution to the input spectrum is small;

3. $\bar{\sigma}(\mathbf{\Lambda}_0)/\underline{\sigma}(\mathbf{\Phi}_u)$ is small, it means that the ratio of external disturbance $\mathbf{r}(k)$ to internal disturbance $\mathbf{e}_0(k)$ is sufficiently high.

Through the above analysis, it is easy to see that $\mathbf{e}_0(k)$ is equivalent to $\xi_i(s)$ in (1), and $\mathbf{r}(k)$ is equivalent to $\mathbf{K}^{-1}(s)\mathbf{K}_h(s)\xi_h(s)$ in (8). Thus, the above conclusion can be applied to load identification. The third requirement states that the ratio of $\xi_h(s)$ to $\xi_i(s)$ should be sufficiently high to obtain a bias free estimation. It appears that this is not satisfied in general if small disturbance data is used, see section 6 for details.

The second requirement states that the effect of feedback should be as small as possible. However, the feedback transfer function can be rather different in actual experiments. Thus, one cannot hope that this requirement is met. In fact, the opposite turns out to be true in all the experiments reported in section 6.

The first requirement compels one to identify a suitable internal disturbance model. This is the only option left, fortunately, this option is found out to be a feasible one.

To summarize, if PEM were to yield satisfactory identification solution using small disturbance data, the experiment has to be *informative* and input data *persistently exciting*. A good internal disturbance model $\mathbf{H}_0$ needs to be captured, next section describes an implementation of PEM to fulfill this requirement.

V. KALMAN FILTERING BASED IMPLEMENTATION

In this section, a Kalman filtering based implementation of PEM is described, it allows the identification model to assume a grey-box structure as is the case in load model identification. The implementation involves five major steps, the details of which are explained below.

1. Obtain $N$ records of measurement data of voltage amplitude $V$, voltage phase angle $\theta$, active power $P$, reactive power $Q$ for the load to be identified. The mean values of the data are removed following a standard argument in system identification [10]. After removing the means, the measurement data are denoted by:

$$\Delta P(k), \Delta Q(k), \Delta V(k), \Delta \theta(k), k = 1,..,N$$

2. Build a suitable load model for identification. The load model is a combination of a static part with constant impedance model and a dynamic part with the third-order induction motor model. The constant current part and constant power part in static part is usually neglected in practice [9]. The load model is linearized; besides, mechanical power of motor is considered as constant. The continuous form of state space load model is as follows:

$$\dot{\mathbf{x}}(t) = \mathbf{A}\mathbf{x}(t) + \mathbf{B}\mathbf{u}(t) + \mathbf{G}\mathbf{e}_0(t)$$
$$\mathbf{y}(t) = \mathbf{C}\mathbf{x}(t) + \mathbf{D}\mathbf{u}(t) + \mathbf{H}\mathbf{e}_0(t) + \mathbf{v}(t) \quad (27)$$

where $\mathbf{u} = [\Delta V\ \Delta \theta]$, $\mathbf{y} = [\Delta P\ \Delta Q]$, and state variables are $\mathbf{x} = [\Delta E_x'\ \Delta E_y'\ \Delta s]$. Here $e_0(t)$ is the internal disturbance, and $\mathbf{v}(t)$ is the measurement noise. The definitions of matrices $\mathbf{A}$, $\mathbf{B}$, $\mathbf{C}$, $\mathbf{D}$ are shown below:

$$\mathbf{A} = \begin{bmatrix} \dfrac{-X}{XT_{d0}'} & s_0 & E_{y0}' \\ -s_0 & \dfrac{-X}{XT_{d0}'} & -E_{x0}' \\ \dfrac{-V_0 \sin(\theta_0)}{XT_j} & \dfrac{V_0 \cos(\theta_0)}{XT_j} & 0 \end{bmatrix}, \mathbf{C} = \begin{bmatrix} \dfrac{V_0 \sin(\theta_0)}{X'} & \dfrac{-V_0 \cos(\theta_0)}{X'} & 0 \\ \dfrac{-V_0 \cos(\theta_0)}{X'} & \dfrac{-V_0 \sin(\theta_0)}{X'} & 0 \end{bmatrix}$$

$$\mathbf{B} = \begin{bmatrix} \dfrac{X-X'}{XT'_{d0}}\cos(\theta_0) & -\dfrac{X-X'}{XT'_{d0}}\sin(\theta_0)V_0 \\ \dfrac{X-X'}{XT'_{d0}}\sin(\theta_0) & \dfrac{X-X'}{XT'_{d0}}\cos(\theta_0)V_0 \\ -\dfrac{E'_{x0}}{XT_j}\sin(\theta_0)+\dfrac{E'_{y0}}{XT_j}\cos(\theta_0) & -\dfrac{E'_{x0}V_0}{XT_j}\cos(\theta_0)-\dfrac{E'_{y0}V_0}{XT_j}\sin(\theta_0) \end{bmatrix}$$

$$\mathbf{D} = \begin{bmatrix} \dfrac{E'_{x0}}{X'}\sin(\theta_0)-\dfrac{E'_{y0}}{X'}\cos(\theta_0)+\dfrac{2P_z^*}{V_0} & \dfrac{E'_{x0}V_0}{X'}\cos(\theta_0)+\dfrac{E'_{y0}V_0}{X'}\sin(\theta_0) \\ \dfrac{2V_0-E'_{x0}\cos(\theta_0)-E'_{y0}\sin(\theta_0)}{X'}+\dfrac{2Q_z^*}{V_0} & \dfrac{E'_{x0}V_0\sin(\theta_0)-E'_{y0}V_0\cos(\theta_0)}{X'} \end{bmatrix}$$

The parameters appearing in **A**, **B**, **C**, **D** are $P_z^*$, $Q_z^*$, $X$, $X'$, $T'_{d0}$, $T_j$, $s_0$, $E'_{x0}$, $E'_{y0}$, $\theta_0$, $V_0$. Since $V_0$ is equal to the mean of $V(k)$ and $\theta(k)$ are simply set to be zero (the load bus is selected as the reference bus), the unknown parameters to be identified include $P_z^*$, $Q_z^*$, $X$, $X'$, $T'_{d0}$, $T_j$, $s_0$, $E'_{x0}$, $E'_{y0}$. Details about the model are referred to [9].

Matrices **G** and **H** describe the influence of internal disturbance, Besides, internal disturbance $\mathbf{e}_0(t)$ and measurement noise $\mathbf{v}(t)$ are usually considered to have zero means, and their covariance matrices are denoted by:

$$\mathcal{Q}=E(e_0(t)e_0(t)^T), \mathcal{R}=E(v(t)v(t)^T), \mathcal{N}=E(e_0(t)v(t)^T)$$

In a usual Kalman filtering implementation, matrices **G**, **H**, $\mathcal{Q}$, $\mathcal{R}$ and $\mathcal{N}$ are determined using domain knowledge. The experience is that, it is generally not difficult to determine matrices **G** and **H**. However, determining the covariance matrices $\mathcal{Q}$, $\mathcal{R}$ and $\mathcal{N}$ takes some fine-tuning efforts. Once matrices **G**, **H**, $\mathcal{Q}$, $\mathcal{R}$ and $\mathcal{N}$ are given, the above optimization problem can be solved using a standard system identification toolbox [17]. A suitable choice of **G**, **H**, $\mathcal{Q}$, $\mathcal{R}$ and $\mathcal{N}$ will be given in simulation and actual system example.

3. Transform the continuous form of state space model (27) into a discrete form as follows:

$$\begin{aligned}\mathbf{x}(k+1) &= \mathbf{A}_d\mathbf{x}(k)+\mathbf{B}_d\mathbf{u}(k)+\mathbf{G}_d\mathbf{e}_0(k) \\ \mathbf{y}(k) &= \mathbf{C}_d\mathbf{x}(k)+\mathbf{D}_d\mathbf{u}(k)+\mathbf{H}_d\mathbf{e}_0(k)+\mathbf{v}(k)\end{aligned} \quad (28)$$

where $\mathbf{A}_d$, $\mathbf{B}_d$, $\mathbf{C}_d$, $\mathbf{D}_d$, $\mathbf{G}_d$, and $\mathbf{H}_d$ are determined following a standard argument in system identification [12].

4. Establish the innovation form as follows:

$$\begin{aligned}\hat{\mathbf{x}}(k+1) &= \mathbf{A}_d\hat{x}(k)+\mathbf{B}_d u(k)+\mathbf{K}_d\mathbf{e}(k) \\ \hat{\mathbf{y}}(k) &= \mathbf{C}_d\hat{x}(k)+\mathbf{D}_d u(k)+\mathbf{e}(k)\end{aligned} \quad (29)$$

where $\mathbf{K}_d$ is given by

$$\mathbf{K}_d = [\mathbf{A}_d\mathbf{P}\mathbf{C}_d^T+\overline{\mathbf{N}}]\times[\mathbf{C}_d\mathbf{P}\mathbf{C}_d^T+\overline{\mathbf{R}}]^{-1}$$

with

$$\overline{\mathbf{N}} = \mathbf{G}_d(\mathcal{Q}\mathbf{H}_d^T+\mathcal{N})$$
$$\overline{\mathbf{R}} = \mathcal{R}+\mathbf{H}_d\mathcal{N}+\mathcal{N}^T\mathbf{H}_d^T+\mathbf{H}_d\mathcal{Q}\mathbf{H}_d^T$$

and **P** is the unique positive definite matrix satisfying the following steady-state discrete Riccati equation [19]:

$$\mathbf{P} = \mathbf{A}_d\mathbf{P}\mathbf{A}_d^T-\mathbf{K}_d(\mathbf{A}_d\mathbf{P}\mathbf{C}_d^T+\overline{\mathbf{N}})^T+\mathbf{G}_d\mathcal{Q}\mathbf{G}_d^T$$

5. Define $\hat{\mathbf{y}}(k)=[\Delta\hat{P}(k)\ \Delta\hat{Q}(k)]$ as the outputs of the identified load model, solve the following optimization problem to determine load model parameters [10, 13]:

$$\min \frac{1}{N}\sum_{k=1}^{N}\|\mathbf{y}(k)-\hat{\mathbf{y}}(k)\|_2^2 \quad (30)$$

where $\hat{y}(k)$ is obtained by:

$$\hat{\mathbf{x}}(k) = (\mathbf{A}_d-\mathbf{K}_d\mathbf{C}_d)\hat{\mathbf{x}}(k)+(\mathbf{B}_d-\mathbf{K}_d\mathbf{D}_d)\mathbf{u}(k-1)+\mathbf{K}_d\mathbf{y}(k-1)$$
$$\hat{\mathbf{y}}(k) = \mathbf{C}_d\hat{\mathbf{x}}(k)+\mathbf{D}_d\mathbf{u}(k)$$

## VI. SIMULATION VERIFICATION

In this section, the 3 machine 9 bus New England test system is used to validate the findings presented. Sampling time in the simulation is equal to 0.01s. The generators are represented by classic second order models [18]. The load at bus 6 is selected as the load to be modelled, loads at other buses are modelled as constant impedances. For ease of explanation, load at bus 6 is an induction machine. Mechanical torque of the machine is regarded as constant and corresponding load parameters are listed in Table 1.

TABLE I.  TRUE PARAMETERS OF INDUCTION MACHINE

| Parameters | $X$ | $X'$ | $T_j$ | $T'_{d0}$ |
|---|---|---|---|---|
| Values | 3.679 | 0.296 | 2 | 0.576 |
| Parameters | $s_0$ | $E'_{x0}$ | $E'_{y0}$ | |
| Values | 0.01 | 0.9014 | -0.1911 | |

In subsequent sections, the identification results of the above parameters are reported. As stated in section 4, to estimate the correct load model using PEM, the internal disturbance model has to be identified accurately. In this simulation, the requirement is validated under the most extreme case: only internal disturbance exists. To simulate it, the mechanical torque of load at bus 6 was injected with white noise (variance 0.002) from 1.5s to 5s.

The relationship (12) will be validated first. Let $[\Delta I_{x6}\ \Delta I_{y6}]$ and $[\Delta V_{x6}\ \Delta V_{y6}]$ denote the current and voltage of the load at 6th bus, thus:

$$\begin{bmatrix}\Delta I_{x6}(s) \\ \Delta I_{y6}(s)\end{bmatrix} = \mathbf{K}(s)\begin{bmatrix}\Delta V_{x6}(s) \\ \Delta V_{y6}(s)\end{bmatrix}, \mathbf{K}(s)=\begin{bmatrix}K_{11}(s) & K_{12}(s) \\ K_{21}(s) & K_{22}(s)\end{bmatrix}$$

where $K(s)$ is an equivalent network matrix defined in (6).

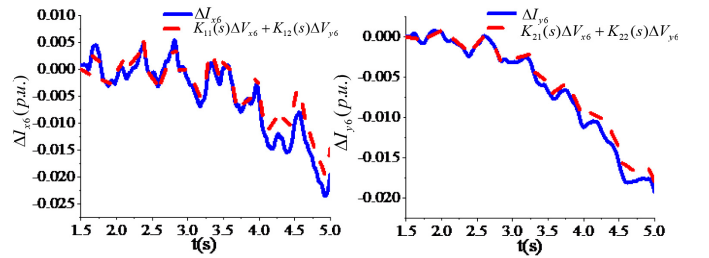

Fig. 4 wave form of $\Delta I_{x6}$, $K_{11}(s)\Delta V_{x6}+K_{12}(s)\Delta V_{y6}$, $\Delta I_{y6}$ and $K_{21}(s)\Delta V_{x6}+K_{22}(s)\Delta V_{y6}$

As can be seen from Fig. 4, the two curves coincide basically which validates the relationship between outputs of load to be identified and inputs to be identified is determined by $K(s)$. The reason for the two curves do not coincide completely is that the load to be identified and the rest component system are nonlinear system. The relationship is derived under the condition that load to be identified and the rest component system are regarded as linear system.

Fortunately, this inaccuracy come from linearization does not prevent us from realizing that this is a closed loop identification.

Next, it is found that the input data is persistently exciting of order greater than 50. To test if the experiment is informative, that is, if $\mathbf{\Phi}_z(\omega) > 0$, first notice that $\mathbf{\Phi}_z$ is Hermitian. It follows that $\mathbf{\Phi}_z$ is positive definite if and only if its eigenvalues are positive [20]. According to our calculation, the ratio of minimum eigenvalue to maximum eigenvalue from 0Hz to 10Hz is positive, this confirms the desired result $\mathbf{\Phi}_z(\omega) > 0$.

In this simulation study, the following internal disturbance model is adopted: $\mathbf{G} = \begin{bmatrix} 0 & 0 & 1/T_j \end{bmatrix}^T, \mathbf{H} = 0$

In order to contrast, an incorrect internal disturbance is also given as follows: $\mathbf{G} = \begin{bmatrix} 0 & 1/T'_{d0} & 0 \end{bmatrix}^T, \mathbf{H} = 0$

The parameter identification results are given in Table 2. The PEM result obtained using suitable internal disturbance is denoted by PEM_A. The result obtained using the incorrect internal disturbance is denoted by PEM_B. The result obtained using the traditional method [4~7] is denoted by TM.

TABLE II. PARAMETERS RESULTS

| Parameters | $X$ | $X'$ | $T_j$ | $T'_{d0}$ |
|---|---|---|---|---|
| PEM_A | 3.2091 | 0.2995 | 1.996 | 0.471 |
| PEM_B | 1.1459 | 0.1877 | 0.1452 | 0.2009 |
| TM | 158.86 | 0.1475 | 2.057 | 45.73 |
| Parameters | $s_0$ | $E'_{x0}$ | $E'_{y0}$ | |
| PEM_A | 0.0103 | 0.9033 | -0.1913 | |
| PEM_B | 0.0168 | 1.0327 | -0.3054 | |
| TM | 0.0109 | 0.8180 | -0.1502 | |

From Table 1 and Table 2, it can be seen that the results of PEM_A are close to the true values. This validates that, when internal disturbance is identified accurately, PEM can produce accurate parameter estimation. However, the results of PEM_B and TM are far away from true values. This validates the importance of modelling the internal disturbance as the bias formulae (26) predicts.

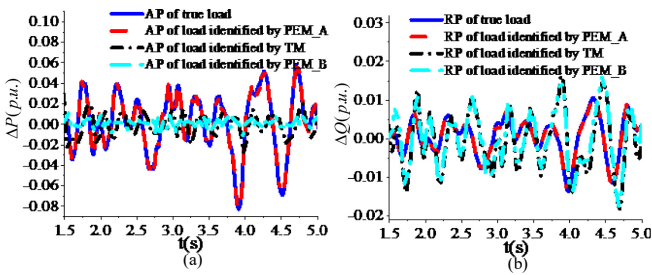

Fig. 5 The reactive and reactive power of the true load, load identified by PEM_A, TM, and PEM_B

To further validate the above observations, the active power (AP) and reactive power (RP) of the true load, the load identified by PEM_A, PEM_B, TM are shown in Fig. 5. It appears that the curves of the true load and the curves of the load identified by PEM_A coincide very well, while the curves of the true load and the curves of the load identified by TM and PEM_B does not fit.

## VII. CONCLUSIONS

The simulations indicate that load model identification using small disturbance data is a closed-loop experiment, and it is feasible to find accurate parameter estimation using the suggested PEM provided that internal disturbance is modeled explicitly. The prerequisites of PEM are discussed in detail, together with some positive empirical findings, it is found that the reported identification effort is useful. Future work should focus on developing a systematic approach to estimate noise covariance matrices $\mathcal{Q}$, $\mathcal{R}$ and $\mathcal{N}$.

REFERENCES

[1] P. W. Sauer. M. A. Pai, *Power System Dynamics and Stability*. NJ, USA: Prentice-Hall, 1999.
[2] W. S. Kao, "The effect of load models on unstable low-frequency oscillation damping in Taipower system experience w/wo power system stabilizers," *IEEE Transactions on Power Systems*, vol. 21, no.2, pp. 463-472, 2001.
[3] D. N. Kosterev, C. W. Taylor, et.al, "Model validation for the August 10,1996 WSCC system outage," *IEEE Trans. Power Syst.*, vol. 14, no. 3, pp. 967–979, 1999.
[4] R. M. He, J. Ma, et.al, "Composite load modeling via measurement approach," *IEEE Trans. Power Syst.*, vol. 21, no. 2, pp. 663–672, 2006.
[5] J. Ma, D. Han, et.al, "Reducing identified parameters of measurement-based composite load model," *IEEE Trans. Power Syst.*, vol. 23, no. 1, pp. 76–83, 2008.
[6] J.K. Kim, K. An, et.al, "Fast and reliable estimation of composite load model parameters using analytical similarity of parameter sensitivity," *IEEE Trans. Power Syst.*, vol. 31, no. 1, pp. 663–671, 2016.
[7] A. Rouhani, A. Abur, et.al, "Real time Dynamic parameter estimation for an exponential dynamic load model," *IEEE Trans. Smart Grid.*, vol. 7, no. 3, pp. 1530–1536, 2016.
[8] F. Viana, Y. Pan, S. Bose, "Bayesian model selection and calibration applied to composite load identification" in *Int. Conf. Transmission and Distribution Conference and Exposition*, Chicago, USA, 14-17 April, 2014, pp. 1-5.
[9] P. Ju, F. Wu, et.al, "Composite load models based on field measurement and their applications in dynamic analysis," *IET Generation Transmission and Distribution.*, vol. 1, no. 5, pp. 724–729, 2007.
[10] V. Michel. V. Vincent, *Filtering and system identification*. UK: Cambridge University Press, 2007.
[11] D. Gan, R. J. Thomas, et.al, "Stability constrained optimal power flow," *IEEE Trans. Power Syst.*, vol. 15, no. 2, pp. 535–540, 2000.
[12] A. K. Tangirala, *Principles of system identification: theory and practice*. USA: CRC Press, 2014.
[13] L. Ljung. *System identification – Theory for the User*. NJ, USA: Prentice-Hall, 1999.
[14] U. Forsell, L. Ljung, "Closed Loop Identification Revisited," *Automatica*, vol. 35, no. 7, pp. 1215–1241, 1999.
[15] A. S. Bazanella, M. Gevers, Mi. Skovi, "Closed Loop Identification of MIMO system: A New Look at Identifiability and Experiment Design," *European Journal of Control*, vol. 16, no. 3, pp. 228–239, 2010.
[16] R. Isermann, M. Munchhof. *Identification of Dynamic Systems*. Berlin, GER: Springer Verlag, 2011.
[17] L. Ljung, System Identification Toolbox Getting Started. [Online].Available: http://ww2.mathworks.cn/help/ident/.
[18] R. D. Zimmerman and C. Murillo-Sanchez, MATPOWER User's Manual. [Online]. Available: http://www.pserc.cornell.edu/matpower/.
[19] A. S. Deshpande, "Bridging a Gap in Applied Kalman Filtering: Estimating Outputs When Measurements Are Correlated with the Process Noise [Focus on Education]," in IEEE *Control Systems Magazine*, vol. 37, no. 3, pp. 87-93, June 2017．
[20] A. van den Bos, *Parameter Estimation for scientist and engineers*, John Wiley & Sons, Inc., 2007.